\documentclass[11pt,epsf]{article}
\usepackage{graphicx}

\begin{document}
\title{Can particles with strength-dependent masses form dark matter of galaxies?}
\author{V. Majernik\\
Institute of Mathematics, Slovak Academy of Sciences, \\
Bratislava, \v Stef\'anikova  47,
 Slovak Republic}

\date{}
\maketitle

\begin{abstract}
The basic idea in this Letter is the assumption that masses of the galactic constituents
(particles of short-living fluctuations) may be functions of strength of the gravitational field.
They may be in galactic space heavier than in the neighborhood of the earth. In the favorable case the total contribution of these constituents
can  be large enough to form the main part of the galactic dark matter.
\end{abstract}



\section{ Introduction}
One of the most important problems in astrophysics concerns the nature
of the dark matter (DM) in galaxies \cite{RU}\cite{EV}\cite{AI}\cite{NO}.
There are many independent lines of evidence that most of the matter  in the universe is
dark, e.g. (i) rotation curves in spiral galaxies, (ii) velocities of galaxies in clusters, (iii)
gravitational lensing, (iv) hot gas in galaxies and clusters, (v) large scale-motion \cite{BE} \cite{CL}.
There exit, in principle, two types of hypotheses for  understanding  the phenomenon of the  galactic DM: the particle  and
field hypotheses. In the particle hypotheses one assumes that DM   consists of the massive  particles occurring in the galactic space.
The prevailing view is that these  particles are  neutral   interacting only gravitationally.
Such are, e.g., the weakly interacting massive particles (WIMPs), additional neutrinos, supersymmetric particles, and a
host of others \cite{BER}.
These particles can be detected experimentally.
Here, the problem is whether they have sufficient large masses in the appropriate abundance at the right localization of a galaxy \cite{GO}.
The field  hypotheses base on the assumptions that DM is created by an interaction of the gravitational field with the galactic  quantum vacuum or particles within a galaxy. For example, there is the assumption that DM is formed by short-lived particles that are created continuously from the quantum fluctuations \cite{ME}
which gain for a very short time a real gravitated mass due to the galactic  field \cite{MA}.
Under  certain circumstances the total mass of these short-living particles
may form a main part of the non-baryonic  matter. Another field hypothesis supposes  that  the gravitationally charged particles  and its antiparticles,  occurring in the galactic space,  form  dipoles whose  interacting with the gravitational field become polarized. The end effect is that the set of polarized dipoles manifest itself as DM \cite{HA} \cite{HA1}\cite{HA2}.
Besides this approach there are attempts to explain the above list of phenomena without existing DM by modifying the
Newtonian law of gravity on large scale \cite{SA}.
Since there is no observational conformation that the gravitational
force falls as $1/r^2$ on large scales the Newton approach to the gravitational
force is modified in that way that it describes, at least, some of the above phenomena.
Many authors pointed out that a modified gravitational force law with the gravitational acceleration $a'$ given by the formula
\begin{equation} \label{1}
a'=\frac{GM_v}{r^2}\left(1+\frac{r}{k}
\right),
\end{equation}
where k represent a constant, could be an alternative to dark matter in galactic halos as an
explanation of the constant-velocity rotation curves of spiral galaxies. However, this
modification of gravitational force do not satisfied the Tully-Fisher law.
Milgrom \cite{MI} proposed an alternative idea, that the separation between the
classical and modified regime  is determined by the value of
the 'critical' gravitational acceleration  rather than the distance scale. He proposed that
\begin{equation} \label{2}
a=\frac{GM_v}{r^2},\qquad a\gg a_c,\qquad a'^2=\left (\frac{aGM}{r^2}\right),\qquad a\ll a_c,
\end{equation}
where the value of the critical acceleration $a_c\approx 8.10^{-8} cm/s^2$ is determined for large
spiral galaxies with $M\sim  10^{44} g$. This equation implies that for $a \ll a_c$ the velocity of galaxies is constant given by
the equation $v^4 = a_0'GM_v$.
This satisfies  the Tully-Fisher law. Although  Milgrom's modification of gravity law
is consisted with a large amount
 of data connected with dark matter it is entirely
{\it ad hoc}.
In what follows, we  determine the amount of  DM in galactic space provided that  masses of galactic constituents  are
direct proportional to strength of the gravitational field whose source is the baryonic matter.

\section {The field of strength-dependent constituents in a galaxy}

If $\rho(r)$ is the density of baryonic matter of a spherically symmetrical galaxy then the strength of its field $E_g(r) $ is
\begin{equation} \label{6}
\nabla  E_g(r)=-G\rho(r).
\end{equation}
Providing that the masses of galactic constituents are directly  proportional to $E_g$
we have
\begin{equation} \label{7}
\nabla E_v =KE_g(r),
\end{equation}
where $E_v(r)$ is  field whose source are the strength-dependent  constituents and $K$ is the proportional constant. The solution of Eq.(4) can be written in the form
\begin{equation} \label{8}
E_v=-\frac{1}{r^2}\int{KE_gr^2} dr.
\end{equation}
For a spherically symmetrical galaxy whose all barionic matter is concentrated in the center, we have
\begin{equation}
E_g(r)=-\frac{GM}{r^2}.
\end{equation}
Inserting $E_g(r)$ into Eq.(4) we have
\begin{equation}\label{9}
\nabla E_v= -\frac{KGM}{r^2},
\end{equation}
where $E_v(r)$ is the field strength due to the constituents  and  $K$ is a constant to be specified. The simplest form  of this constant $K$, containing the mass of galaxy
$M$ gravitational constant $G$ and Milgrom's critical constant $a_c$, appears to be the expression
$$ K=\sqrt{\frac{a_c}{GM}}.$$
where $a_c\sim 10^{-8} cms^{-2}$. It is worth to remark that, according to Massa \cite{MAS}
there are several reason to think that the acceleration
$a_c=10^{-8} cms^{-2}$ plays a fundamental role in the cosmology : (i) $a_c$ is the surface gravity of a pion
(mass $ \sim 10^{-25} gr$, radius $\sim10^{-13} cm$), (ii) the surface gravity of a protostar (mass $\sim 10^{34} gr$, radius $\sim10^{17} cm$),
(iii) the surface gravity of a typical spiral galaxy  (mass $ \sim 10^{44} gr$, radius $\sim10^{22} cm$), (iv) the surface gravity of the observable universe (mass $ \sim 10^{56} gr$, radius $\sim10^{28} cm$). Coincidentally, the value of said acceleration is roughly equal to
$a_c=10^{-8} cms^{-2}$, where according to  Milgrom, the Newton law is substituted by a modified gravitational law. Eq.(7) has
the solution
\begin{equation} \label{10}
E_v(r)=-\frac{\sqrt{a_cGM}}{r}.
\end{equation}

Since numerically $a_c\approx G$ and  $\sqrt{a_cG}\approx G$ we can approximately write  $K\approx 1/\sqrt{M}$
\begin{equation} \label{11}
\nabla E_v(r)=-\frac{\sqrt{a_cGM}}
{r}\approx \frac{G\sqrt{M}}{r}.
\end{equation}

The total strength of galactic field $E_t(r)$ is the sum of $E_g(r)$ and $E_v(r)$
\begin{equation}\label{7a}
E_t(r)=E_g(r)+E_v(r),
\end{equation}
where
\begin{equation} \label{7b}
E_t(r)=-\frac{GM}{r^2} \qquad E_2=-\frac{\sqrt{a_cGM}}{r}.
\end{equation}
$E_t(r)$  represents the modified  gravity law. The first term expresses the familiar field due to barionic matter while the second term gives the field
whose source are  the quantum fluctuations
or enlarged masses of particles in the gravitational field.
The distance $r_e$ where $E_g$ and $E_v$ are equal is
\begin{equation} \label{z}
r_e=\sqrt{\frac{GM}{a}} \approx \sqrt{M}
\end{equation}
so we have
\begin{equation} \label{11b}
E_g\approx -\frac{GM}{r^2}\left(1+\frac{r}{\sqrt{M}}\right)
\end{equation}
Near the galactic center the field  $E_g$ prevails while for $r >\sqrt{a_c}GM$ the field $E_v(r)$ is dominated.
Next we are looking for effective  mass $m_e$ of galaxy whose gravitational field in $r$ is identical with $E_t$ if no dark matter would exists,
i.e. when
\begin{equation}\label{12}
\frac{GM}{r^2}\left(1+\frac{r}{\sqrt{M}}\right)=\frac{Gm}{r^2}.
\end{equation}
From Eq.(14) it follows

\begin{equation}     \label{33}
 m_e\approx M(1+\frac{r}{\sqrt{M}}).
\end{equation}
For example, the effective mass $m_e$ of a typical spiral galaxy with mass $M\approx 10^{44} g$ and the radius $r \approx 10^{22}cm$
is, according  Eq.(15), equal to $2\times M $.

\section{Two-term gravitational law}

As said above Milgrom suggested that Newton's second law of motion should be reexamined in the case of galactic motions.
His basic idea was that at very low accelerations, corresponding to large distances, the second law broke down.  To make it work better,
he added a new constant into Newton law called the critical acceleration $a_c$. Milgrom's \cite{MII} modified Newtonian dynamics (MOND)
has been critiqued mainly in two point \cite{NO}.
Why this modification works by galactic but not in earth gravitational field which are sometimes much smaller then field at the edge of galaxies and why the transition between both is so {\it abrupt}. As shown in previous chapter
we modified the gravitational law which represents a continuous transition between the  regime near galaxy center and that on the edge of galaxy in the form
\begin{equation} \label{q}
F_m(r)=-\frac{GM}{r^2}-\frac{\sqrt{a_cGM}}{r}.
\end{equation}
Let us suppose that $F_m(r)$ is valid generally  for all gravitating systems and one may it apply
to the gravitating system 	
outside the galaxies. For example, to the solar system.
The two-term  law for the  solar system has the form

$$ F_\odot=-\frac{GM_\odot}{r_g^2}-\frac{\sqrt{a_cGM_\odot}}{r_g},$$

 where $M_\odot\approx 2.10^{30} kg$ is the mass of the sun and $r_g=AU\approx 1.5.10^{11} m$ is the distance between the earth and sun.
 Now, we can find whether the field $T_2$  has observable effect in distance $r_g$.
  Inserting these values  into Eq.(12) we  can determine the distance $r_e$ where $T_1=T_2$,
Doing this we find  $$r_e=\sqrt{\frac{GM_\odot}{r_c}}\approx 10^{15} m.$$
Inserting $r_e$ into Eq.(16) we get
  $$T_1=\frac{GM_\odot}{r_e^2}=6.2.10^{-3} m sec^{-2} $$
and
$$T_2=\frac{\sqrt{a_cGM_\odot}}{r_e}=\frac{7.10^{-11}\sqrt{2.10^{30}}}{1.5.10^{11}}= 6.5.10^{-7}m sec^{-2}.$$ The ratio  $T_1/T_2\approx 10^{4}$, i.e the field $T_1$ in $r_e$  is much stronger then the field $T_2$. Hence, $T_2$ is in distance $r_g=AU$ virtually negligible. Only 	
behind the distance $r_e= 10^{15} m$, i.e.
far from the solar system,  $T_2$ prevails $T_1$.

From what has been said so far it follows: (i) the assumption that  masses of certain galactic constitutions  are strength-dependent combining with Milgrom's critical acceleration, build in the proportional constant $K$, leads to the two-term gravity law.
The first term of this law expresses the familiar force while the second term gives the force
whose source are the constituents generated by galactic  field itself.
The two-term law guarantees that the field given by first term of Eq.(16) is dominant in the neighborhood of the galactic center  while  at the  edge of galaxy prevails the second term. (ii) The two-term law is smooth in contrast with the original Milgrom's law which is abrupt. (iii)
The effective mass of a galaxy is linear function of galactic radius. (iv) The two-term law may be verified in the conditions of solar system.
For example, through the influence of second term on  the orbits of planets in solar system.

Acknowledgement

The work was supported by the Slovak Academy Grant Agency VEGA-2/0059/12.

\end{document}